\documentclass[twocolumn]{aastex631}

\newcommand{\bs}{\boldsymbol}

\newcommand{\R}{\mathcal{R}}
\newcommand{\T}{\mathcal{T}}
\newcommand{\Msun}{M_\odot}

\usepackage{graphicx,grffile,ulem,url,threeparttable,xcolor,soul,booktabs,natbib,amsmath,mathrsfs}
\usepackage{txfonts,rotating,longtable, needspace,subfigure,multirow,array}

\shorttitle{Baryonic Effects - Spin Reconstruction}
\shortauthors{Sheng et al.}

\begin{document}

\title{Baryonic Effects on Lagrangian Clustering and Angular Momentum Reconstruction}

\correspondingauthor{Hao-Ran Yu}
\email{haoran@xmu.edu.cn}

\author[0000-0002-9891-338X]{Ming-Jie Sheng}
\affiliation{Department of Astronomy, Xiamen University, Xiamen, Fujian 361005, China}

\author[0000-0001-5277-4882]{Hao-Ran Yu}
\affiliation{Department of Astronomy, Xiamen University, Xiamen, Fujian 361005, China}

\author[0000-0003-4813-8482]{Sijia Li}
\affiliation{Department of Astronomy, Xiamen University, Xiamen, Fujian 361005, China}

\author[0000-0001-7075-6098]{Shihong Liao}
\affiliation{Department of Physics, University of Helsinki, Gustaf H\"allstr\"omin katu 2, 
FI-00014 Helsinki, Finland}
\affiliation{Key Laboratory for Computational Astrophysics, National Astronomical Observatories,
Chinese Academy of Sciences, Beijing 100101, China}

\author[0000-0001-9953-0359]{Min Du}
\affiliation{Department of Astronomy, Xiamen University, Xiamen, Fujian 361005, China}

\author[0000-0001-8913-626X]{Yunchong Wang}
\affiliation{Department of Physics, Stanford University, 382 Via Pueblo Mall, Stanford, CA 94305, USA}
\affiliation{Kavli Institute for Particle Astrophysics and Cosmology, P. O. Box 2450, Stanford University, 
Stanford, CA 94305, USA}

\author[0000-0003-2504-3835]{Peng Wang}
\affiliation{Shanghai Astronomical Observatory, 80 Nandan Road, Shanghai 200030, China}

\author[0000-0002-7697-3306]{Kun Xu}
\affiliation{Department of Astronomy, School of Physics and Astronomy, Shanghai Jiao Tong University, Shanghai 
200240, China}

\author[0000-0002-3185-1540]{Shy Genel}
\affiliation{Center for Computational Astrophysics, Flatiron Institute, 162 Fifth Avenue, New York, NY 10010, USA}
\affiliation{Columbia Astrophysics Laboratory, Columbia University, 550 West 120th Street, New York, NY 10027, USA}

\author[0000-0003-2946-8080]{Dimitrios Irodotou}
\affiliation{Department of Physics, University of Helsinki, Gustaf H\"allstr\"omin katu 2, 
FI-00014 Helsinki, Finland}

\begin{abstract}
Recent studies illustrate the correlation between the angular momenta of cosmic structures and their Lagrangian properties. 
However, only baryons are observable and it is unclear whether they reliably trace the cosmic angular momenta.
We study the Lagrangian mass distribution, spin correlation, and predictability of dark matter, gas, and stellar components of 
galaxy--halo systems using IllustrisTNG, and show that the primordial segregations between components are typically small. 
Their protoshapes are also similar in terms of the statistics of moment of inertia tensors.
Under the common gravitational potential they are expected to exert the same tidal torque
and the strong spin correlations are not destroyed by the nonlinear evolution and complicated
baryonic effects, as confirmed by the high-resolution hydrodynamic simulations.
We further show that their late-time angular momenta traced by total gas, stars, or the central
galaxies, can be reliably reconstructed by the initial perturbations. These results suggest that baryonic 
angular momenta can potentially be used in reconstructing the parameters and models related to the 
initial perturbations.
\end{abstract}


\section{Introduction} \label{sec.intro}
The large-scale structure (LSS) of the universe is primarily driven by the dynamics of dark matter (DM).
After recombination, baryonic matter decouples from radiation and follows the clustering of DM under 
gravity. Hence, the matter distribution on a large scale can be probed by various of tracers, such as galaxies,
resulting in rich cosmological information \citep[e.g.,][]{1969ApJ...155..393P,2005MNRAS.360L..82R,2021JCAP...06..024M}.
At low redshifts, the nonlinear structure formation generates vorticities in the matter velocity field. 
This nondecaying small-scale vector mode actually reflects primordial density perturbation on larger
scales, via the tidal torque theory \citep{1970Afz.....6..581D,1984ApJ...286...38W}. In particular,
the tidal environments of the protohalos in the Lagrangian space, characterized by the Hessian of the primordial 
gravitational potential, torque those protohalos in a persistent way such that the virialized DM
halos at low redshifts tend to keep the predicted angular momentum directions \citep{2002MNRAS.332..325P}
and magnitudes \citep{2021PhRvD.103f3522W}.
Thus, their angular momenta provide independent cosmological information,
including, e.g., the reconstruction of primordial density and tidal fields
\citep{2000ApJ...532L...5L,2001ApJ...555..106L},
the effects of cosmic neutrino mass \citep{2019PhRvD..99l3532Y,2020ApJ...898L..27L} and
dark energy \citep{2020ApJ...902...22L}, possible detection of chiral
violation \citep{2020PhRvL.124j1302Y,2022PhRvD.105h3512M},
and the understanding of galaxy intrinsic alignments \citep[e.g.,][]{2001MNRAS.320L...7CF, 
2011JCAP...05..010B,2015JCAP...10..032S,2018MNRAS.473.1562W}.

Unfortunately, unlike the mass of DM halos that can be inferred by gravitational lensing,
the rotation of DM halos are difficult to observe, and one can only expect the angular momenta
of galaxies or other baryonic tracers to be the proxies of that of DM halos. 
The three-dimensional (3D) spins (hereafter we refer to ``angular momentum direction'' as ``spin'' for brevity)
of galaxies are readily observable \citep[see the discussions in][]{2019ApJ...886..133I,2021NatAs...5..283M}.
This parity-odd observable is free from the contamination of linear perturbation theory, and 
many approaches are trying to understand it observationally or theoretically.
Most recently, \citet{2020PhRvL.124j1302Y} proposed the idea of predicting the spin mode of protohalos
by using the $E$-mode clustering in Lagrangian space, referred to as ``spin reconstruction,'' 
and by using this method \citet{2021NatAs...5..283M} for the first time discover a weak but significant 
correlation between the observational galaxy spins and the reconstructed cosmic initial conditions.

The importance of this correlation deserve further explanation and investigation. 
First, the reconstructed initial conditions, given by ELUCID \citep{2014ApJ...794...94W,2016ApJ...831..164W},
use only galaxy positions without their spins, so the correlation demonstrates that the spins of cosmic structures,
even traced by baryons, indeed contain additional cosmological information beside galaxy/halo locations.
Second, the correlation is found in Lagrangian space, where Fourier modes are still linear and directly
related to cosmological constraints.
Third, this observational attempt involves both known and unknown physical processes and systematic errors. 
Regarding to the last point, the errors include those in the reconstructed initial conditions,
in the Lagrangian space remapping (S. Li et al. 2022, in preparation), and in the complicated observations 
of galaxy spins, etc. While these techniques continue to be improved,
the underlying gravitational and baryonic processes are yet to be studied separately. 
In particular, it is still unclear whether and how baryonic components trace the DM 
across the cosmic evolution and galaxy formation.
The baryonic effects include gas cooling, star and galaxy formation, and supernova and black hole feedbacks,
which are highly nonlinear and their effects on baryonic angular momenta cannot be modeled by
cosmological perturbation theories. 
In this work, we use the state-of-the-art magnetohydrodynamical (MHD) simulations IllustrisTNG 
\citep{2018MNRAS.480.5113M,2018MNRAS.475..624N,
2019MNRAS.490.3234N,2018MNRAS.475..648P,2019MNRAS.490.3196P,2018MNRAS.475..676S,2019ComAC...6....2N} 
to study these baryonic effects on the angular momentum generation, and quantify how these primordial spin modes 
can be traced by the baryonic matter at low redshifts.

In the rest of this article, Sec.\ref{sec.meth} describes the simulation and basic analyses.
Sec.\ref{sec.resu} shows the spin conservation and reconstruction results for DM and baryonic components.
The conclusion and discussion are presented in Sec.\ref{sec.conclu}.

\begin{table}[htbp]
  \caption{Counts of Halos and Mean Galaxy Counts per Halo (Galaxies with Stellar Mass 
  Threshold $M_{\rm s}=10^{9}\,h^{-1} \Msun$)
  in Different Mass Bins 
  (Mass Units: $h^{-1} \Msun$)}\label{Tab1}
  \resizebox{8.6cm}{!}{
  \begin{tabular}{p{2cm}<{\centering}cc}
  \hline
  \hline
  Halo Mass &Halo Counts & Mean Galaxy Counts per Halo\\
  \hline
  $[10^{11.5},10^{12})$    &$2956$    &$1.2$\\
  $[10^{12},10^{12.5})$    &$1051$    &$2.1$\\
  $[10^{12.5},10^{13})$    &$355$     &$4.9$\\
  $[10^{13},10^{13.5})$    &$124$     &$13.2$\\
  $[10^{13.5},+\infty)$    &$60$      &$63.6$\\
  \hline
  \end{tabular}}
\end{table}

\begin{figure*}[htbp]
  \centering    
  \subfigure[\;FOF $\#9$, $M_{\rm tot}=1.7\times10^{14}\,h^{-1} \Msun$] {    
  \includegraphics[width=1\columnwidth]{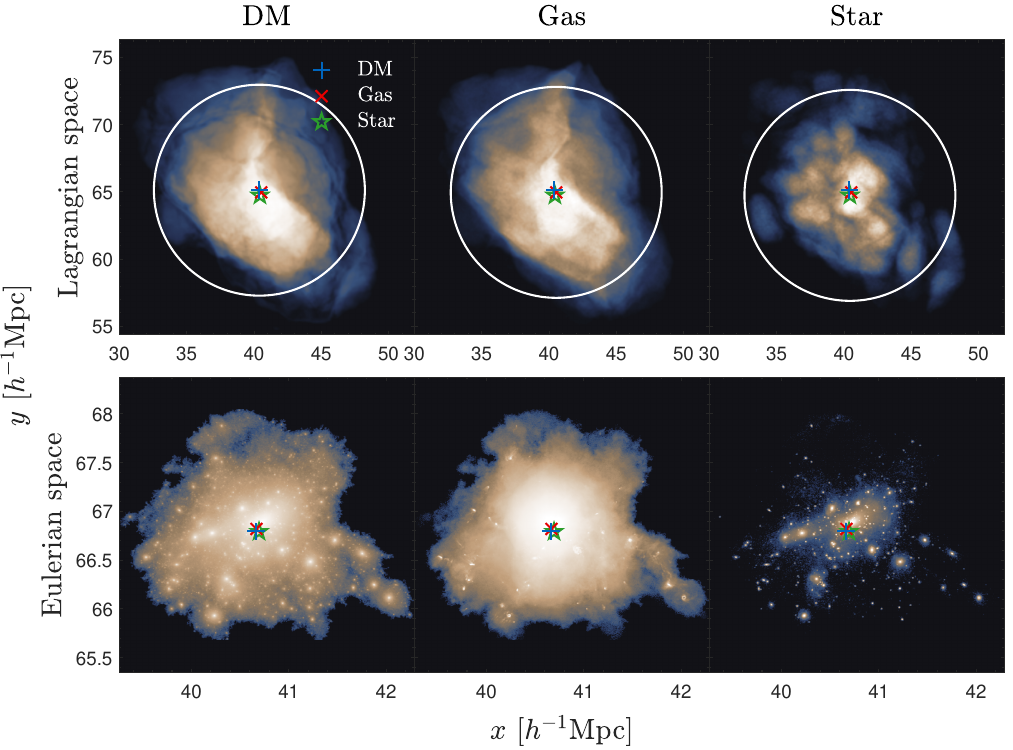}  
  }     
  \subfigure[\;FOF $\#163$, $M_{\rm tot}=1.1\times10^{13}\,h^{-1} \Msun$] { 
  \includegraphics[width=1\columnwidth]{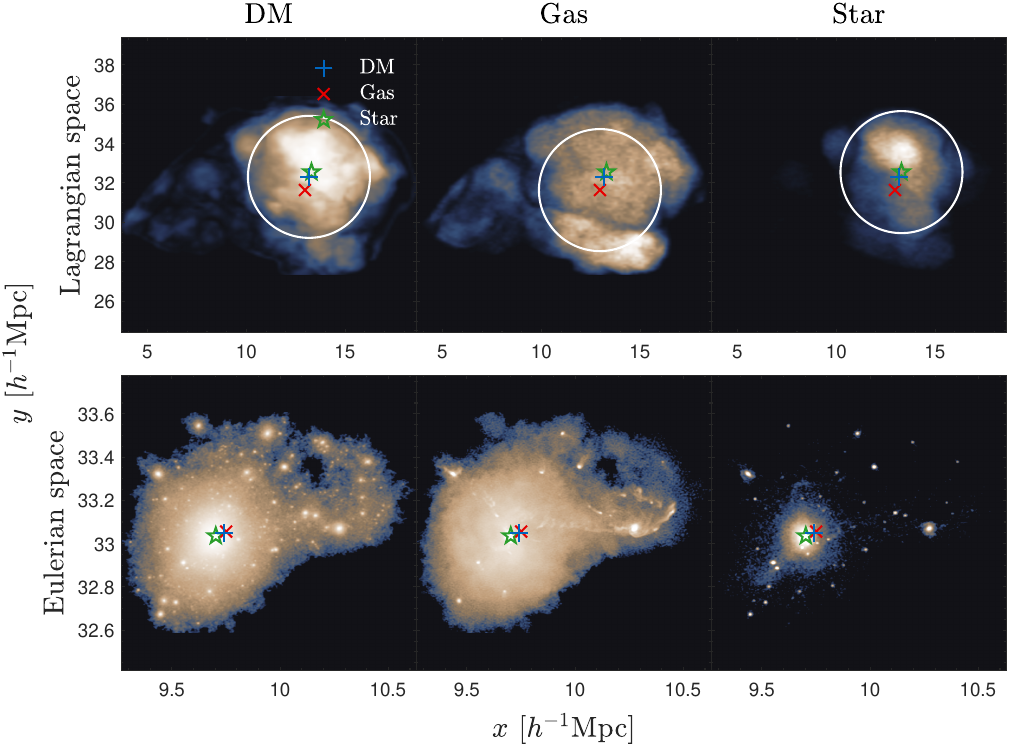}     
  }     
  \subfigure[\;FOF $\#1322$, $M_{\rm tot}=1.3\times10^{12}\,h^{-1} \Msun$] {    
    \includegraphics[width=1\columnwidth]{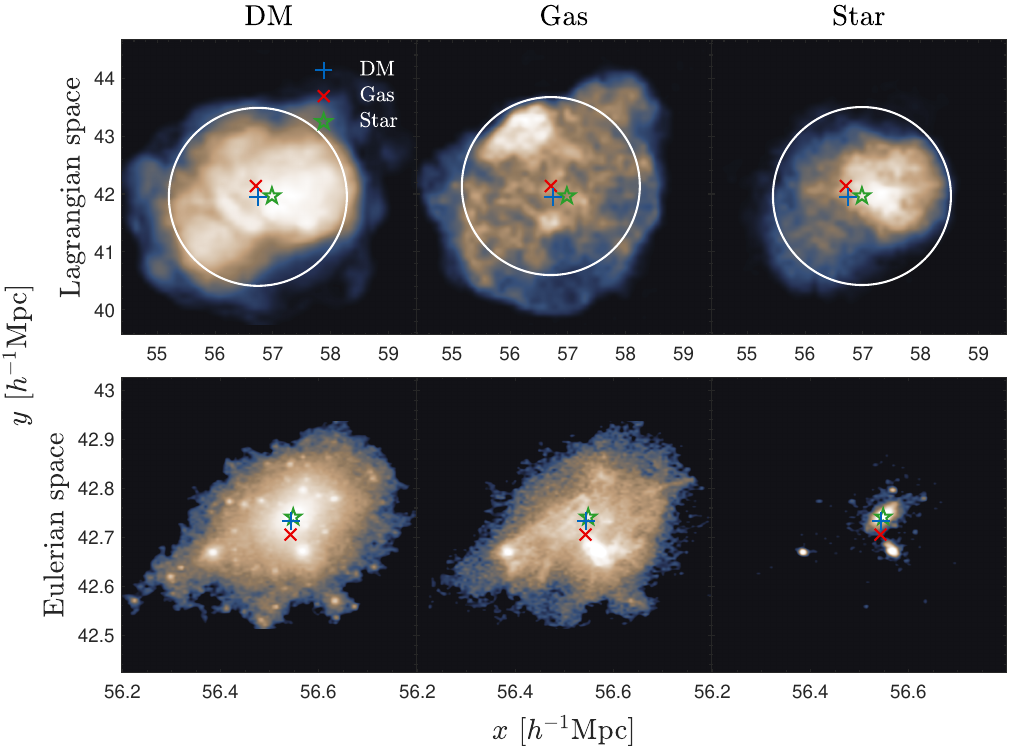}  
  }
  \subfigure[\;FOF $\#3760$, $M_{\rm tot}=3.9\times10^{11}\,h^{-1} \Msun$] {    
    \includegraphics[width=1\columnwidth]{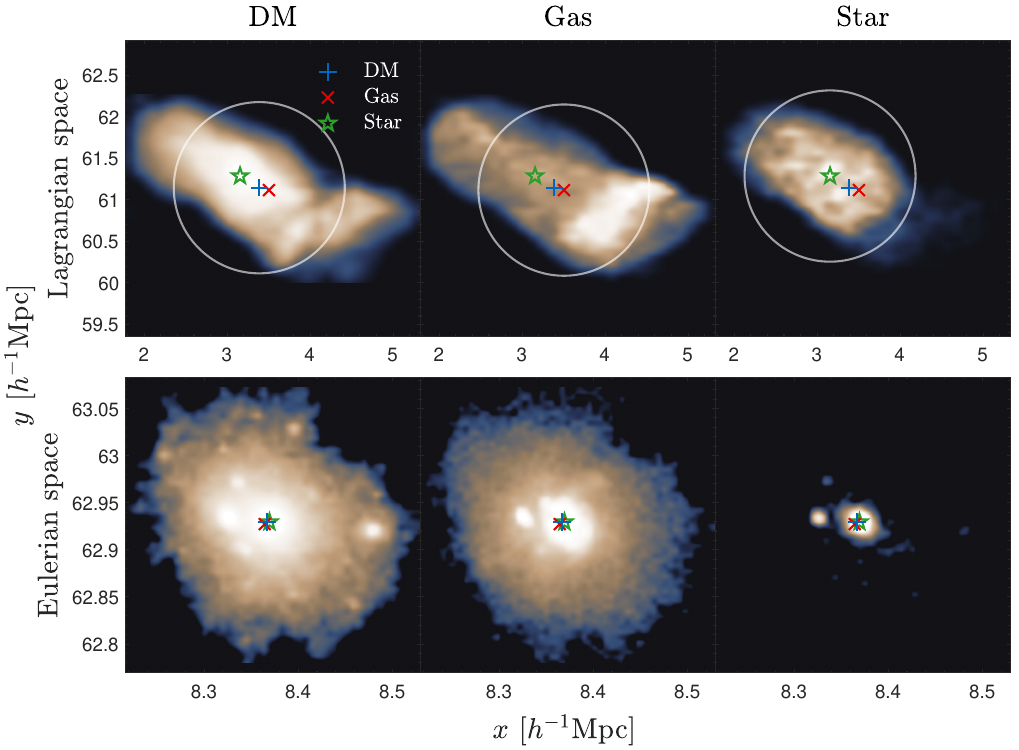}  
  }    
  \caption{Projected column density of DM, gas, and stellar components in Lagrangian and Eulerian spaces. 
  Their CoM are marked by respective symbols. The radii of circles in Lagrangian space indicate $r_q$.}     
  \label{fig.cd}     
\end{figure*}

\begin{figure}[htbp]
  \centering
  \includegraphics[width=0.95\linewidth]{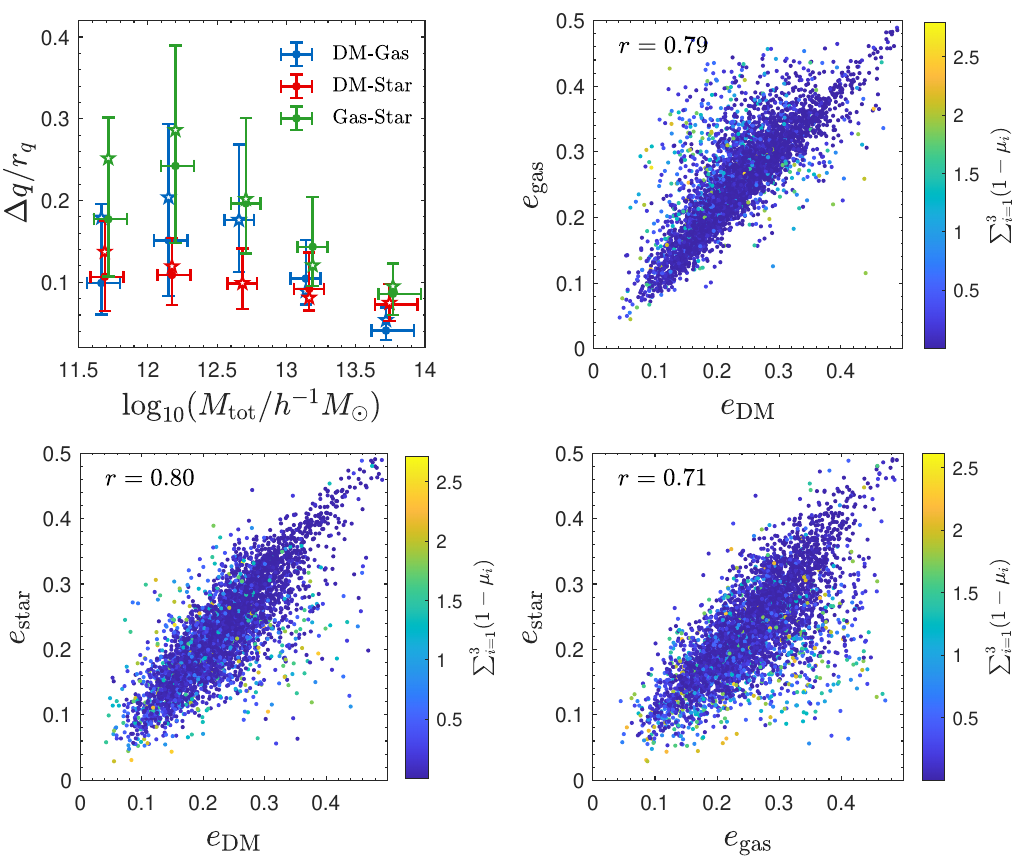}
   \caption{Quantification of the mass distributions of DM, gas, and stellar components in Lagrangian space. 
   The top left panel shows the distributions of normalized CoM offsets $\Delta q/r_q$ for five halo mass bins. 
   The center, lower/upper boundaries of the error bar represent the median, $25\%$/$75\%$ quartiles of the distribution. 
   Note that the distribution is not Gaussian and thus not symmetric; `$\star$' symbols indicate the expectation values of the
   distributions. The remaining three panels show the alignments of Lagrangian distributions. We use ellipticity to characterize the shape. Points are colored according to the overall axes alignment offsets $\sum_{i=1}^{3}(1-\mu_i)$.
   }
   \label{fig.shape}
\end{figure}

\section{Methodology and Basic Analyses}\label{sec.meth}
The IllustrisTNG simulations are a suite of MHD galaxy formation simulations using the {\small AREPO} code \citep{2010MNRAS.401..791S,2020ApJS..248...32W}. 
In this study, the main results are given by the TNG100-1 simulation, which starts with $1820^3$ DM particles and $1820^3$ gas 
cells in a periodic cubic box with a comoving length $75 \,h^{-1} {\rm Mpc}$ per side.
The initial condition is generated with the {\small N-GENIC} code \citep{2005Natur.435..629S} by perturbing a 
``glass'' particle load \citep{1996clss.conf..349W} with the Zel'dovich approximation. The adopted 
cosmological parameters are from the Plank 2015 results \citep{2016A&A...594A..13P}, i.e., $\Omega_{\rm m}=
0.3089$, $\Omega_{\rm b}=0.0486$, $\Omega_\Lambda=0.6911$, and $h=0.6774$. The mass resolutions for DM particles and gas cells are $m_{\rm DM}=5.1\times10^6\,h^{-1} \Msun$ and $m_{\rm gas}=
9.4\times10^5\,h^{-1} \Msun$ (on average), respectively.

DM halos and subhalos are identified with friends-of-friends (FOF; \citealp{1985ApJ...292..371D}) and 
{\small SUBFIND} algorithms \citep{2001MNRAS.328..726S}.
The halo total mass $M_{\rm tot}$ is defined as the sum of the individual mass of every particle/cell, of all types
in consideration, i.e., DM, gas, and stars. 
We only consider FOF halos with total mass $M_{\rm tot}\geq 10^{11.5}\,h^{-1} \Msun$,
yielding a halo catalog contains $4546$ samples with particle IDs, positions, velocities, and other astrophysical
properties for DM, gas, and stellar components. Shown in Table.\ref{Tab1}, the total samples
are divided into five mass bins and list halo counts in each mass bin.  

To study the primordial spin mode, we need to trace the halo mass elements back to Lagrangian space.
For DM, we can simply trace them by following the particle IDs. For gas cells and star particles, 
we trace their tracer particles \citep{2013MNRAS.435.1426G} back to the initial condition. 
The TNG100-1 simulation contains $2\times1820^3$ tracer particles. 
To see the distribution of each component of halos and corresponding protohalos more intuitively, 
in Figure \ref{fig.cd}, we randomly select four halos in four different mass bins of Table.\ref{Tab1}
and plot the column density projected onto the $x$--$y$ plane. 
Subpanels correspond to different components in Lagrangian (initial condition) and 
Eulerian (redshift $z=0$) spaces, respectively. The halo masses, halo IDs, and coordinates in TNG100-1 are explicitly shown in the figure.

The mass distributions in Lagrangian space play important roles in the angular momentum production,
in which the Lagrangian space volume occupation (size), center-of-mass (CoM) location, and 3D shape are key
ingredients. The Lagrangian size is directly related to the total halo mass; the CoM location can be reliably 
reconstructed (S. Li et al. 2022, in preparation); and the 3D shape can be characterized, up to the second moment (quadrupole), 
by the moment of inertia tensor. Since the structure formation inside each DM halo is mostly virialized, we expect, and 
it is verified by simulation that, the above properties for different components are similar. 

Focusing on the sizes first, despite the mass distributions of DM and gas are quite diffuse in Eulerian space while 
on the contrary for stars, their Lagrangian sizes are all comparable to the equivalent protohalo radius 
$r_q$, indicated by radii of circles in Figure \ref{fig.cd}. Here $r_q$ is defined as
\begin{equation}
  r_q\equiv \left(\frac{2M_{\rm tot}G}{\Omega_{\rm m}H^{2}_{0}}\right)^{1/3},
\end{equation}
where $G$ is the Newton's constant and $H_0$ is the Hubble's constant.

Next, we quantify CoM offsets. In Lagrangian and Eulerian spaces, the $+,\times,\star$ symbols represent CoM positions of 
DM, gas, and stellar components respectively. Due to the baryonic effects, the Lagrangian--Eulerian mappings of DM 
and baryons are not exactly the same \citep{2017MNRAS.470.2262L} and result in these CoM offsets. 
The top left panel of Figure \ref{fig.shape} shows the distribution 
of CoM offsets $\Delta q$ normalized by $r_q$, for each halo mass bin. The Lagrangian CoM offsets of different components 
are typically small compared to their physical sizes. Especially, for more-massive halos, both the offset and the deviation
become smaller. This mass dependence could be explained by the fact that a deeper gravitational potential is more 
capable of locking baryons, and thus the system is less affected by the environment, and consequently different 
components are more likely to originate from the same Lagrangian region.

The moment of inertia tensor $I_{jk}=\sum_i m_i {x}'_{j}{x}'_{k}$ describes the shape of a mass distribution 
up to quadrupole, where $m_i$ is the particle mass, and ${\bs x}'$ is the particle position 
relative to CoM. The eigen-decomposition of $I_{jk}$ gives the the primary, intermediate, and minor 
axes of the mass distribution and their spatial alignments. The eigenvalues are sorted as 
$\lambda_1>\lambda_2>\lambda_3$, associated with the eigenvectors ${\bs V}_1$, ${\bs V}_2$, ${\bs V}_3$.
We use 3D ellipticity
\begin{equation}
  e=\frac{\lambda_1-\lambda_3}{2(\lambda_1+\lambda_2+\lambda_3)},
\end{equation}
and the alignment eigenvectors to characterize the shape and alignment correlations.
For the latter, the alignments are quantified by the cosine $\mu_i$ ($i=1,2,3$) of the acute angle between 
${\bs V}_i$ of the components in comparison. In the top right panel and bottom panels of Figure \ref{fig.shape}, 
we plot the correlation of ellipticity between different components, with overall axes alignment 
offsets $\sum_{i=1}^{3}(1-\mu_i)$ indicated by colors. We can see that, for each halo--galaxy system, the Lagrangian
counterparts for three components generally have similar shapes and spatial alignments.
%Note that some mathematical special cases will slightly reduce the statistical correlation, e.g., when two components 
%in comparison have similar axes lengths and $\lambda_2 \simeq \lambda_3$, while $\mu_2$, $\mu_3\simeq0$, there will 
%appear to be a huge axes alignment offsets, but in fact they are perfectly aligned.
The Pearson correlation coefficients $r(e_{\rm DM},e_{\rm gas})$, $r(e_{\rm DM},e_{\rm star})$, $r(e_{\rm gas},e_{\rm star})$ are shown in the three subpanels.

\begin{figure}[htbp]
  \centering
  \includegraphics[width=0.85\linewidth]{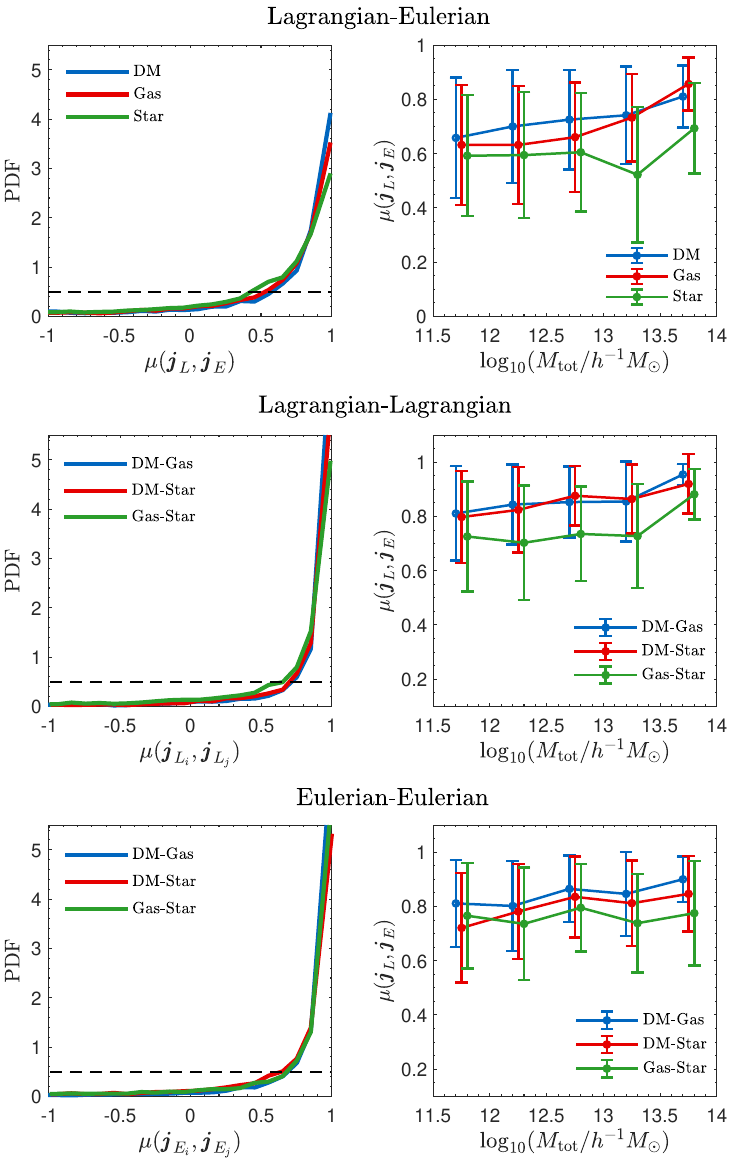}
   \caption{Conservation of spins of DM, gas and stellar components through 
   the cosmic evolution (top) and alignment of spins between these components in Lagrangian (middle)
   and Eulerian (bottom) spaces. The left panels show the PDFs of $\mu(\bs j_L,\bs j_E)$, $\mu(\bs j_{L_i},\bs j_{L_j})$ and 
   $\mu(\bs j_{E_i},\bs j_{E_j})$, where $i,j=$ DM, gas, and stellar components in colored solid lines, 
   with black dotted line indicates the PDF of a random distribution. The right panels show the average $\mu$ 
   as a function of halo total mass binned following Table.\ref{Tab1}. Error bars represent the standard deviation of 
   the distributions.} 
   \label{fig.cc}
\end{figure}

\section{Results}\label{sec.resu}
\subsection{Alignment and Conservation of Spins}
In this section we start with a study on the spin properties of DM and baryons of halos.
In Lagrangian (Eulerian) space, the angular momentum vector $\bs j_L$ ($\bs j_E$) of a certain component 
(e.g., DM, gas, or stars) is computed as
\begin{eqnarray} 
  \bs j_L&=&\sum_i m_i (\bs q_i - \bar{\bs q})\times (\bs u_i - \bar{\bs u}),\\
  \bs j_E&=&\sum_i m_i (\bs x_i - \bar{\bs x})\times (\bs v_i - \bar{\bs v}),
\end{eqnarray}
where $m_i$, $\bs q_i$ ($\bs x_i$) and $\bs u_i$ ($\bs v_i$) are the particle mass, Lagrangian (Eulerian) position, 
and velocity of the $i{\rm th}$ particle, while $\bar{\bs q}$ ($\bar{\bs x}$) 
and $\bar{\bs u}$ ($\bar{\bs v}$) are the Lagrangian (Eulerian) CoM position and mean velocity of this 
component.

We use the cosine of the angle between two vectors $\bs j_L$ and $\bs j_E$ to quantify the 
cross-correlation between their directions,
\begin{equation}
  \mu(\bs j_L,\bs j_E) \equiv \frac{\bs j_L \cdot \bs j_E} {|\,\bs j_L\,|\,|\,\bs j_E\,|} \in [-1,1].
\end{equation}
Randomly distributed 3D vectors results in a top-hat distribution of $\mu$ with $\langle\mu\rangle=0$. 
The top left panel of Figure \ref{fig.cc} shows
the probability density functions (PDFs) of $\mu(\bs j_L,\bs j_E)$ for DM, gas, and stellar components of all 
samples. 
The expectation values $\langle\mu\rangle$ take $0.68$, $0.64$, and $0.60$, respectively, and the PDFs of 
$\mu(\bs j_L,\bs j_E)$ obviously depart from a top-hat distribution, suggesting that $\bs j_L,\bs j_E$ 
directions for each component are strongly correlated. Note that comparing to the DM component,
the gas and stellar components have experienced a series of baryonic processes \citep[e.g.,][]{2017ApJ...841...16D},
especially the stellar and AGN feedback \citep[e.g.,][]{2017MNRAS.466.1625Z}, but they still remain a strong correlation. 
This indicates that the memory of the initial tidal fields of the baryonic components is not fully erased by the 
baryonic processes. In addition, the stellar component shows a weaker 
correlation compared to gas, which could be explained by the fact that the galaxy stellar spins are affected by galaxy 
merger events \citep[e.g.,][]{2022ApJ...936..119L} and star formation processes. 
In the middle left and bottom left panels of Figure \ref{fig.cc}, we show the 
correlations of spins between different components in Lagrangian and Eulerian spaces, respectively. 
The middle left panel suggests that the spins for each component of protohalos are strongly correlated,
which originates from the similar mass distributions of these components and the same tidal torque they feel 
in Lagrangian space, as shown in Figure \ref{fig.shape}. 
In addition, the gas and stellar components show weaker 
correlations, which is consistent with the result of Figure \ref{fig.shape} that these two components 
show a larger Lagrangian CoM offset and weaker similarity in shape. 
The bottom left panel shows that these alignments are well conserved through the cosmic evolution.
Considering the halo mass dependence of these correlations, the right panels in Figure \ref{fig.cc} show that, 
more-massive halos have a better conservation of spins for each component through the cosmic evolution, 
and they also show a stronger alignment of spins between different components.

\begin{figure}[htbp]
  \centering
  \includegraphics[width=0.85\linewidth]{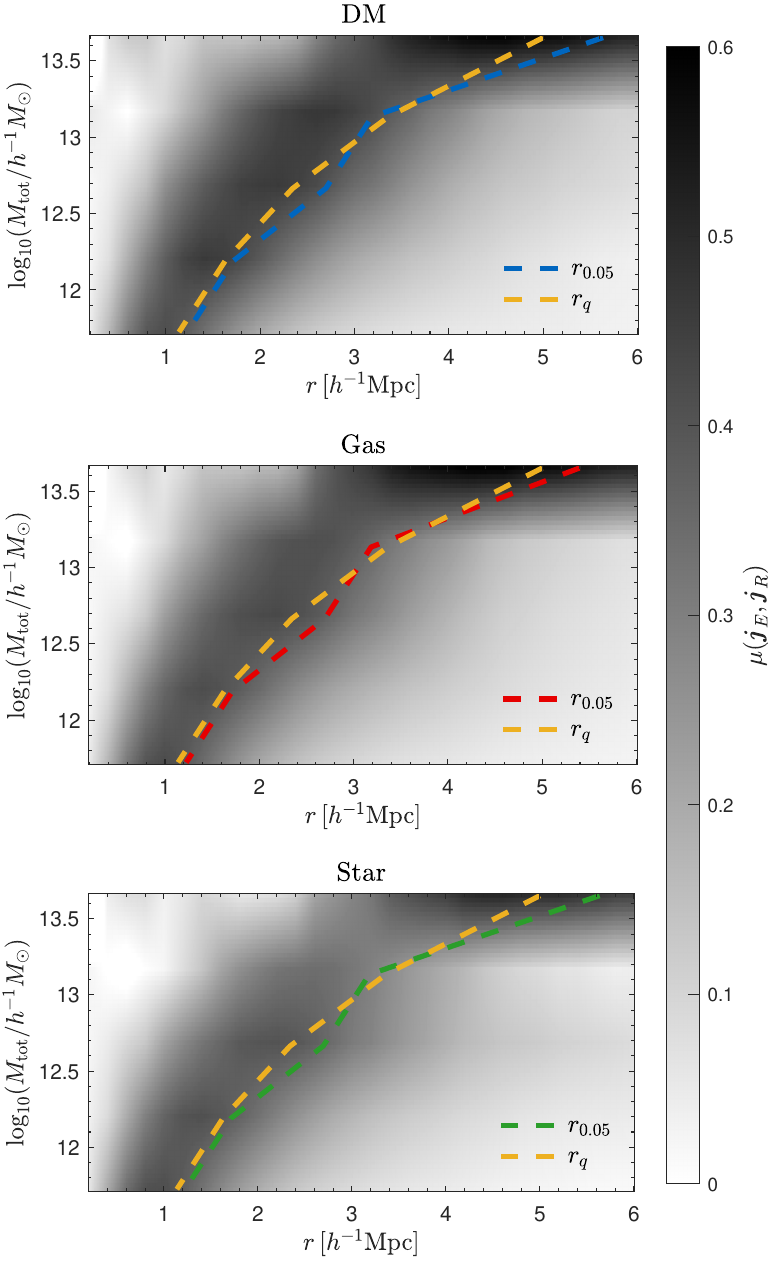}
   \caption{The cross-correlation coefficients between Eulerian ${\bs j}_E$ halo spins and spins
   reconstructed using Eq.(\ref{eq.rec}) from known initial conditions ${\bs j}_R$ for DM, gas, and stellar 
   components. Darker colors show better reconstruction.
   Optimal $r_{0.05}$ (blue, red, and green dashed curves) and the Lagrangian equivalent radius $r_q$ (yellow dashed curves)
   of the protohalo are also plotted.}
   \label{fig.rec}
\end{figure}

\subsection{Spin Reconstruction}
In this section we reconstruct the spins for baryonic components based on their 
Lagrangian space properties analogous to the spin reconstruction of DM component.
In the tidal torque theory, the 
initial angular momentum vector of a protohalo that initially occupies Lagrangian volume $V_L$ is 
approximately by 
\begin{equation}
  j_\alpha \propto \epsilon_{\alpha\beta\gamma}I_{\beta\kappa}T_{\kappa\gamma},
\end{equation}
where $\bs{I}=(I_{\beta\kappa})$ is the moment of inertia tensor of $V_L$, $\bs{T}=
(T_{\kappa\gamma})$ is the tidal tensor acting on $\bs{I}$, and $\epsilon_{\alpha\beta\gamma}$ 
is the 3D Levi-Civita symbol.
%collecting the antisymmetric components generated by the misalignment between $\mathbf{I}$ and $\mathbf{T}$, and the Einstein summation convention is assumed in this article. 
%As the inertial tensor $I_{\beta\kappa}$ is closely aligned with the tidal tensor $T_{\kappa\gamma}$ 
%\citep[e.g.][]{2000ApJ...532L...5L,2001ApJ...555..106L,2002MNRAS.332..325P}, 
\citet{2020PhRvL.124j1302Y} propose the spin-reconstruction method for halo spins as
\begin{equation}
  {\bs j}_R=(j_\alpha)_R \propto \epsilon_{\alpha\beta\gamma} \bs\T_{\beta\kappa} \bs\T^{+}_{\kappa
  \gamma},
  \label{eq.rec}
\end{equation}
where $\bs \T_{\beta\kappa}$ and $\bs\T^{+}_{\kappa\gamma}$ are tidal fields constructed as the Hessian of
the initial gravitational potential smoothed at two different scales $r,R$. 

We use the {\small N-GENIC} code to obtain the gravitational potential field $\phi$ of the 
TNG100-1 initial condition, and convolve $\phi$ with a Gaussian window function
to obtain the smoothed potential field for calculating $\bs \T_{\beta\kappa}$ and $\bs\T^{+}_{\kappa\gamma}$.
The actual calculation is done in Fourier space \citep{2020PhRvL.124j1302Y}.
Here we characterize the smoothing scale of the Gaussian window function by defining the 
radius $r_{0.05}$ at which the window function drops to $5\%$ of its maximum.

\begin{figure}[htbp]
  \centering
  \includegraphics[width=0.80\linewidth]{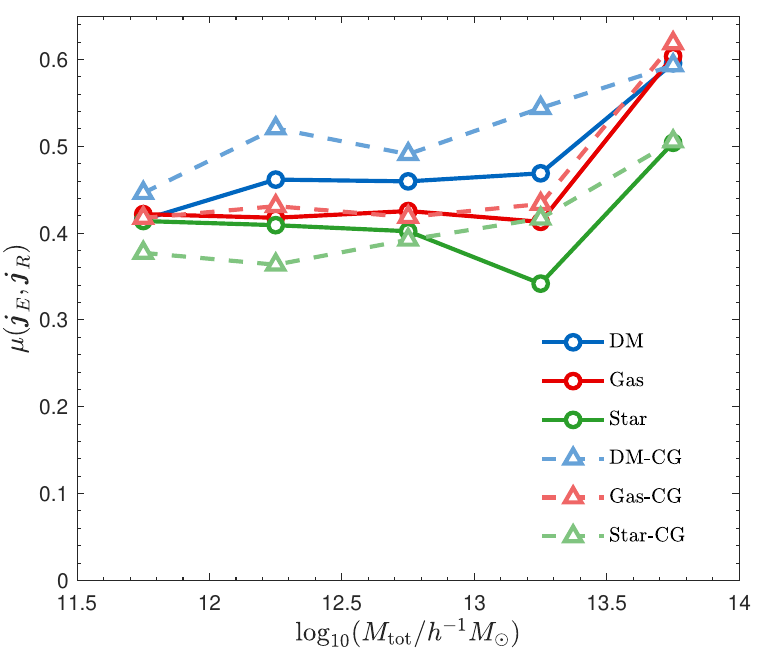}
   \caption{Maximally achievable cross-correlation coefficients as a function of halo total mass for components of 
   total halos (solid line) and central galaxies (dashed line). The results for DM, gas, and stellar 
   components are plotted in blue, red, and green, respectively.}
   \label{fig.rm}
\end{figure}

We cross-correlate the spins ${\bs j}_E$ of each halo component in Eulerian space with the spins ${\bs j}_R$ 
reconstructed by applying Eq.(\ref{eq.rec}).
We follow \citet{2020PhRvL.124j1302Y}, which suggests that by choosing $R \to r_{+}$ the cross-correlation 
maximizes. In Figure \ref{fig.rec}, from top to bottom, we plot the cross-correlation 
coefficients $\mu({\bs j}_E,{\bs j}_R)$ for DM, gas, and stellar components and the optimal $r_{0.05}$ 
which maximize $\mu$ in different mass bins.
We find that, for different components, the optimal $r_{0.05}$ are fairly similar and close to the equivalent 
protohalo radius $r_q$ in Lagrangian space. 
%Besides, the optimal $r_{0.05}$ increase from $\sim 1\,h^{-1} {\rm Mpc}$ for Milky Way-sized galaxies to $\sim 5\,h^{-1} {\rm Mpc}$ 
%for galaxy groups/clusters, agreeing with the results from $N$-body simulation in \citet{2020PhRvL.124j1302Y}.

In Figure \ref{fig.rm}, we plot the maximally achievable cross-correlation coefficients as a function of halo total 
mass for DM, gas, and stellar components of total halos.
The DM component has a maximum cross-correlation between 0.4 and 0.6, while slightly lower for gas and stellar 
components. 
This discrepancy is similar to Figure \ref{fig.cc} in that the gas and stellar components 
have experienced a series of baryonic processes.
But overall they still have a maximally achievable cross-correlation of $0.35$ to $0.5$, which opens up the possibility 
of using the spins of gas and stellar components to constrain the cosmic initial conditions.

The DM halos in the sample could contain more than one galaxy, shown in Table.\ref{Tab1}.
They include one central galaxy and possibly several satellite galaxies. 
The angular momentum of such a galaxy--halo system includes individual galaxy spin 
angular momenta $\sum{\bs j}_{\rm spin}$ and orbit angular momentum ${\bs j}_{\rm orb}$. 
We further compute the cross-correlation between the spin angular momenta of the central galaxy
${\bs j}_{\rm spin, CG}$ of each halo and the reconstructed halo spins. In Figure \ref{fig.rm}, by repeating the similar 
analyses for the components of total halo, we obtain a similar statistical correlation for central galaxies.

\section{Conclusion and Discussion}\label{sec.conclu}
We study the conservation and predictability of the spins of DM, gas, and 
stellar components of galaxy--halo systems in TNG100-1 simulation. We conclude the following:
  \begin{itemize}
    \item The mass distributions of the DM, gas, and stellar components of galaxy--halo systems in Lagrangian
     space are quite similar, in terms of locations, sizes, and shapes. Some offsets exist but are typically small 
     compared to their physical sizes. 
    \item Similar mass distributions between DM and baryonic components lead to a strong spin correlation between
     them in Lagrangian space, which is mostly conserved across the cosmic evolution.
     Besides, the spins of baryonic components between halos and their protohalos are also very well correlated, 
     similar to that of the DM component. The memory of the initial perturbations of these components is not fully erased by 
     the nonlinear structure formation.
     \item Similar Lagrangian space mass distributions also enable us to use a universal spin-reconstruction algorithm
     for different components: similar locations, sizes, and shapes result in similar reconstructing locations, 
     smoothing scales, and results, respectively. The spins of DM and baryonic components of total halos can be predicted 
     by this method in Lagrangian space. In addition, a similar result exists for central
     galaxies. This provides us with the possibility of using observable galaxy spins to constrain the cosmic initial conditions.
  \end{itemize}

For a convergence test on resolutions, we also test the result 
performed on a lower resolution simulation TNG100-3. We find a good convergence between different resolution simulations 
with the same statistical results. Besides, various of baryonic and galaxy formation models may result in different 
spin correlations. We notice that some other studies using different hydrodynamic simulations all confirm DM and 
baryon spin correlation in Eulerian space \citep[e.g.,][]{2015ApJ...812...29T,2019MNRAS.488.4801J}. 
While many studies based on $N$-body simulations confirm the strong spin correlation for DM component 
between the Lagrangian and Eulerian spaces \citep[e.g.,][]{2002MNRAS.332..325P,2021PhRvD.103f3522W},
we can qualitatively infer that using different galaxy formation 
models will not significantly affect our statistical results. Comparing the results quantitatively based on 
IllustrisTNG and other hydrodynamical simulations are left to future works.

One of the main challenges of using baryonic angular momenta to reconstruct the initial conditions is to obtain the precise 
observational data of galaxy spin.
For massive halos, various kinds of techniques are able to observe ${\bs j}_{\rm spin, CG}$, ${\bs j}_{\rm orb}$,
or the overall angular momentum of diffuse gas in the halo \citep[e.g.,][]{2008MNRAS.389.1179L,2017MNRAS.467.1965H,2018MNRAS.477.4711G}.
From our initial investigation, they are all reliable tracers of halo spins and primordial spin modes.
Their discrepancies might contain DM--baryon segregation information \citep{2017MNRAS.470.2262L}.
Another challenge is the reconstructing methods. The existing reconstructing methods, such as ELUCID, use only galaxy positions 
without their spins. Whether additional spin information can help improving reconstructing the initial conditions is worth studying.

On larger, more linear scales, cosmic filament spins provide more cosmological information complimentary to 
that of galaxies and halos \citep{2022PhRvD.105f3540S}. It would be interesting to study the cosmic filament spins
traced by baryonic matter and the spin correlations in galaxy--filament systems.

\section*{acknowledgements}
We thank the anonymous referee for valuable suggestions.
This work is supported by National Science Foundation of China Grants No. 11903021 and No. 12173030. SL acknowledges 
the NSFC grant No. 11903043. SL and DI acknowledge the support by the European Research Council via ERC Consolidator 
Grant KETJU (No. 818930).
The authors acknowledge the support by the China Manned Space Program through its Space Application System.
The Flatiron Institute is supported by the Simons Foundation.
The IllustrisTNG simulations were run on the HazelHen Cray
XC40 supercomputer at the High Performance Computing Center Stuttgart 
(HLRS) as part of project GCS-ILLU of the Gauss
Centre for Supercomputing (GCS).

\bibliography{mingjie_ref}{}
\bibliographystyle{aasjournal}

\end{document}